# A Conceptual Framework for Understanding Empathy in Physics faculty

Alia Hamdan, Ash Bista, Dina Newman, Scott Franklin

Abstract:
Physics, like many scientific disciplines, has long struggled with attracting and retaining a diverse population and fostering inclusivity. While there have been improvements in addressing equity issues within physics, significant challenges remain. Faculty members play a crucial role as change agents in promoting cultural transformation within academic environments. Empathy, a fundamental component of effective teaching, mentoring, and collegiality, is essential for fostering a student-centered and holistic approach in academia. However, understanding how empathy functions within the specific context of physics, including its interaction with power dynamics and other contextual factors, remains underexplored. This study presents a theoretical model of empathy development among physics faculty as they engage with students and colleagues. Conducted at a private R2 institution, the study involved four rounds of interviews from summer 2023 to spring 2024. The initial two rounds (summer 2023) included eight participants, followed by 19 participants in fall 2023, and nine participants chose to follow up with us in spring 2024. All participants were physics faculty members, either in teaching or tenure-track positions. The developed model builds on previous research by introducing new complexities in the understanding of empathy. It identifies key mediators, including reflective witnessing, personal experiences, and empathetic concern, as well as moderators such as individual experience, emotions, motivation, values, and situational information. The model delineates both cognitive and affective pathways of empathy, providing a nuanced framework for understanding how empathy develops and influences faculty interactions in the physics discipline.

I.  Introduction

*"I have a student that's you know, we were planning on working on whatever project today, but this student needs to just sit and talk for an hour about some stuff that they have to unload, … that's the priority. I wasn't expecting that but that's what we need to do today, not whatever their project is"*

Empathy, the ability to understand and share the feelings of another person, has emerged as a critical concept within the field of education. Nevertheless, there is an ambiguity in how researchers operationalize and define it and a disproportionate focus on cognitive (intellectual understanding) and emotional (a sharing of emotions) empathy, with less work on compassionate empathy (the desire to alleviate another's suffering) ((Håkansson Eklund & Summer Meranius, 2021). Physics often frames itself as an "objective" discipline, (Soler, 2015) devaluing or ignoring emotional and social aspects, and consisting of a "culture of no culture"(Traweek,

1992). This creates barriers for people from diverse backgrounds, and hinders the inclusivity of physics as a discipline. Addressing this requires a subjective lens, recognizing how individuals approach the field, listening and learning from the insights and experiences of others. In short, an *empathetic* approach to those within and on the fringes of the community is needed.

The importance of empathy in the classroom has been recognized by the National Academy (*Education for Life and Work*, 2012), with an operational definition of (Meyers et al., 2019) "the degree to which instructors work to deeply understand students' personal and social situations, feel caring and concern in response to students' positive and negative emotions, and communicate their understanding and caring to students through their behavior." Fostering empathy builds strong relationships between students and instructors (Lunn et al., 2022), and instructors who demonstrate empathy help students develop a stronger sense of social responsibility and ethical decision-making skills (Strobel et al., 2013), positively impact student achievement (Postolache, 2020), and garner greater respect from their students (Feshbach & Feshbach, 2009). Empathy has been studied as a transferable skill, with numerous interventions developed to enhance student empathy in the workspace (Algra & Johnston, 2015; David Carlson & Dobson, 2020; Hess et al., 2017; McCurdy et al., 2020; Walther et al., 2016, 2020).

A variety of psychometric instruments to measure empathy (Baron-Cohen & Wheelwright, 2004; Davis, 1983; Spreng et al., 2009) have been developed and used in academic contexts. The Toronto Empathy Questionnaire (TEQ) (Spreng et al., 2009) measures empathy as a single factor defined as "an emotional process, or an accurate affective insight into the feeling of another." The Empathy Quotient (EQ) (Baron-Cohen & Wheelwright, 2004) was developed to diagnose people with high functioning autism and defines empathy as a multidimensional construct that includes cognitive empathy, emotional reactivity, and social skills. The Interpersonal Reactivity Index (IRI) is the most widely used instrument and treats empathy as a multidimensional construct (Davis, 1983; Davis & others, 1980) with four subscales: perspective taking (PT), fantasy scale (FS), empathetic concern (EC) and personal distress (PD). Using both the TEQ and a newly developed Instructor Empathy Practices questionnaire (IEP), Ross et al. (2023) found that students' empathy scores were significantly correlated with Grade Point Average, student level classification, and race. Students perceived instructors to be more empathetic when actively listening (nodding, eye contact) and showing flexibility and patience. Perceptions of instructor empathy led to students reporting that they worked harder, felt less stressed and had higher motivation. There is some pushback against psychometric instruments, questioning whether they are measuring empathy or other emotions (Baron-Cohen & Wheelwright, 2004).

Research on the lack of diversity in physic**s** has identified several factors that suggest empathy as a critical underlying issue. Power dynamics, informal social networks, and gatekeeping practices (Bonifazi et al., 2022; Hodari et al., 2022c, 2022a, 2022b) all indicate that individuals in positions of privilege may be unaware of their own unconscious biases or may prioritize a perceived meritocratic system over the lived experiences of others. Even well-intentioned physics faculty may struggle to recognize or address student challenges(Dancy & Hodari, 2023), leading to a significant disconnect between students' experiences and faculty perceptions.

This paper focuses on physics faculty due to the unique and problematic "culture of no culture" within the field (Traweek, 1992). Despite decades of research and efforts, physics remains a highly homogeneous and exclusive discipline. The cultural norms, often implicit and difficult to identify, contribute to an "out of sight, out of mind" mentality, leading to increasing numbers of individuals feeling marginalized. We approach this disconnect as a disruption in the empathetic pathway, presenting a novel framework that connects noticing, empathy, and action.This paper builds on previous the previous study (Merrill et al., 2024a) that outlines a theoretical model on how physics faculty develop and use empathy in their work with students. This research expands on previous research to gain a deeper understanding on the differences between both cognitive and affective empathetic experiences by investigating other hidden factors that contribute to the pathways. This study uses qualitative analysis and data collection methods to ground the work in physics faculty experiences and aims to facilitate research on student-faculty interactions, culture of physics, faculty action and decision making, and development of interventions that work on making physics a more inclusive space.

II.     Literature review: the empathetic pathway

Prior work on empathy has compartmentalized the construct into three main subparts, namely cognitive empathy, affective empathy and concern.

Zaki and Ochsner (2012) identify mechanisms through which empathy can develop and connect them to specific neural processes. Affective empathy or "experience sharing" involves vicariously sharing a target's internal states, whereas cognitive empathy or "mentalizing" is the process by which one explicitly considers targets' states and their sources. These can give rise to a "prosocial concern," defined as a motivation to improve targets' experiences. Experience sharing is often linked to 'neural resonance,' with the same neural system activating in response to a personal experience and the observation of another in the similar experience (c.f. Rizzolatti & Sinigaglia, 2010, Keysers et al., 2010 and Lamm & Singer, 2010). In contrast, cognitive empathy activates a network of midline and superior temporal structures associated with 'self-projection' (Buckner & Carroll, 2007), enabling individuals to represent states beyond their immediate 'here and now,' including future, past, hypothetical scenarios, and others' perspectives. Zaki and Ochsner highlight the limitations of relying solely on neuroscientific data to explore empathy and call for research in more complex realistic settings through the lens of social models.

Yu and Chou (2018) build on LeDoux's (1998) work on categorizing emotions as high and low route, to conceptualize a dual route model of empathy, differentiating between the cognitive and affective processing of emotions. Affective empathy-- "one's emotional, sensorimotor, and visceral response to the affective state of others"--is automatic and fast (low route), and triggered by the 'Mirror Neuron System', where the brain copies an emotion inferred from others (Iacoboni, 2009; Rizzolatti & Sinigaglia, 2016). Cognitive empathy–"the ability to understand or explicitly reason the subjective mental states, perspective, or intention of others"-- is a slower, more complex process (high route) that requires one to formulate theories of what another person is experiencing, thus requiring a high cognitive effort. They also include a factor that introduces the action component, termed "prosocial behavior," which is defined as taking action to address the state or needs of another person.

Building on these prior works, we developed a simplified empirical *empathetic pathway* (Figure 1) from interviews with physics faculty (Merrill et al., 2024a). Using a constructivist grounded theory approach, common themes and patterns were identified on how empathy is perceived and expressed, and its impacts on teaching and faculty-student interactions. In addition to the basic pathway, specific characteristics and mediators were identified. Shared lived experience mediate paths between affective empathy, attention and noticing, and empathetic action, while contextual information acts as a mediator for pathways between cognitive empathy, attention/noticing and empathetic action. Merrill et al. looked at the individual as the basic unit of analysis, but this work suggested that we think of empathy as a dynamic construct construed through communication. This paper builds on the prior frameworks to identify mediator, and moderator relations along the pathways and more fully develop a cohesive empathetic framework.

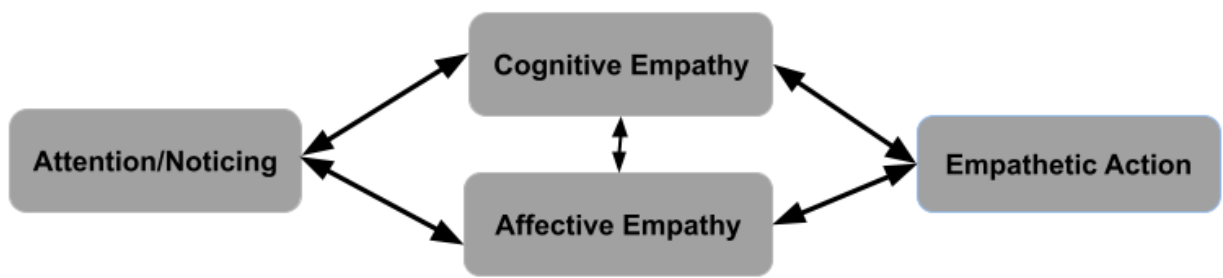

Figure 1. Simplified version of the empathetic pathway, as first illustrated in (Merrill et al., 2024b). Dual arrows are used to show multi-directionality of pathways. A dual arrow is also placed between cognitive and affective empathy to signify the ability to move between paths and towards empathetic action.

III. Data Collection & Analysis

Physics faculty at a private R2 institution were interviewed in four rounds across a period spanning summer 2023 to spring 2024. The first two rounds (summer 2023) were conducted by RM (Rachel Merrill) and included a total of eight participants; AH (Alia Hamdan) and AB (Ash

Bista) subsequently interviewed 19 participants in fall 2023 (round 3) and 9 participants in fall 2023 and spring 2024 (round 4). All participants were physics faculty in either teaching or tenure track positions. The same participants were interviewed in rounds 1 and 2; participants from round 4 were a sample of those in round 3. Participants were asked to choose their own pseudonyms and pronouns, so we did not utilize a structured naming convention. We employed a semi-structured interview format, allowing for open-ended questions while maintaining a focus on the research topic. This approach encouraged participants to elaborate on their experiences and perspectives, and not all questions were mandatory to answer.

 In the interview, we delved into various aspects of faculty experiences and viewpoints. We started by asking the physics faculty to recount situations where they noticed a student struggling, how they defined empathy, and whether its application varied depending on the context, requesting specific examples to substantiate their responses (round 1,3). We also explored what faculty considered as empathy towards themselves by asking for instances when they felt particularly understood by students or colleagues and how they perceived others' understanding of them. In follow-up interviews (round 2,4), we sought more detailed information on each aspect, specifically asking for examples where others clearly demonstrated cognitive or affective empathy. To gain insight into communication practices, we requested faculty to outline their communication methods with students and colleagues, including instances where communication might have failed or been misunderstood. Additionally, we inquired about their perceptions of their own roles and responsibilities to understand their expectations as faculty members. Overall, the interview protocol aimed to provide a reflective space for participants to consider their thoughts and feelings about their experiences and interactions. The interviews ranged from 30 minutes to 1 hour in length and were audio-recorded. Audio recordings were transcribed using Otter.ai, reviewed for accuracy and sent back to the participants for verification, where no changes were made.

Data analysis employed a constructivist grounded theory approach (Mills et al., 2006) which followed an iterative process. First, detailed memos were written after each interview and served as a bridge between the raw data and the emerging themes, summarizing key points and highlighting important aspects of the discussions. Next, researchers reviewed the memos and transcripts to identify recurring themes and patterns. This involved an interplay between the individual memos and the broader dataset, allowing for the emergence of key concepts. Once themes were identified, transcripts were coded, with codes attached to segments of text that captured relevant concepts or ideas. The codebook was a dynamic document, evolving as new themes emerged and existing codes were refined. Once the codebook was finalized, all transcripts were re-coded to establish and strengthen the internal validity. The initial codes were organized according to the four main factors identified by Merrill et al. (2024): attention/noticing, cognitive empathy, affective empathy, and empathetic actions. For instance, cognitive empathy included codes like inferring others' thoughts and emotions and recognizing differences between oneself and others.

An iterative thematic analysis was then conducted on the coded data. This process involved identifying new codes related to the development of empathy and empathetic actions, grouping these codes into broader thematic categories, and pinpointing mediating factors in how physics faculty conceptualize empathy. Codes were categorized as either mediators or moderators based on their impact on the pathway and their prevalence in examples. Mediators were factors present in the majority of cases, influencing the pathway significantly, while moderators were factors that influenced the pathway but were not essential.This ensured that the codes were being applied consistently across the entire dataset. For an inter-rater reliability (IRR) RM and AH each independently coded interview segments (1 interview from the initial 7) and compared results to identify any inconsistencies in the application of codes. Disagreements were resolved through discussion, leading to a robust understanding of the data. A second round occurred with AH and AB, each independently coding segments from each interview (round 3,4) and comparing results. Disagreements were resolved through discussion and reform of the codebook. Final results were presented to the authors as well as a wider interdisciplinary group (CASTLE) for confirmation of plausibility, and collaborative discussions held to determine effective visualizations. Member checking involved sending completed drafts of each presentation, poster, and written document (such as conference proceedings and journal articles) containing information from participants' interviews. One participant raised concerns about the potential identifiability of a narrative. After discussing the issue, we decided to make adjustments to the details to protect the identities of both the faculty member and the students involved.

IV. Results

We add complexity to the empathetic pathway model by identifying mediators and moderators (Field-Fote, 2019) that affect different paths. Mediators explain how independent variables influence dependent variables while moderators affect the strength or direction of those relationships. Below we initially separately discuss the two primary pathways (cognitive and affective empathy) and then, once each has been presented, discuss how the two are connected. The results will first describe the reflective pathway with cognitive empathy, detailing and explaining the mediators—reflective witnessing and empathetic concern—as well as the moderators, which include experience, emotions, motivations and values, and situational information, illustrated in figure 2. The following subsection will explore the reflexive empathetic pathway involving affective empathy, elucidating the identified mediators—experience and empathetic concern—and the moderators—emotions, motivation and values, and situational information—along with how these factors were discerned from the data and depicted in figure 3. Subsequently, figure 4 depicts the combined theoretical model outlining how physics faculty develop either cognitive or affective empathy and subsequently engage in (or refrain from) empathetic actions.

**Developing a theoretical framework; Drawing from data**

    A. Cognitive empathy pathway

The majority of narratives provided by the participants of this study, physics faculty, center on cognitive empathy, demonstrating their understanding and perceptions of the perspectives of others. Figure 2 illustrates the reflective empathetic pathway, identifying reflective witnessing as a crucial mediating factor. This process involves utilizing background information gathered through communication and attention to various aspects to develop cognitive empathy. The diagram also shows experience as a moderator that can either enhance or impede the effectiveness of reflective witnessing, and personal emotions as a moderator in shaping cognitive understanding of students. Additionally, the figure depicts the pathway to action once understanding is achieved, emphasizing the roles of empathetic concern, personal motivations and values, and situational information as key factors influencing decisions to take empathetic actions.

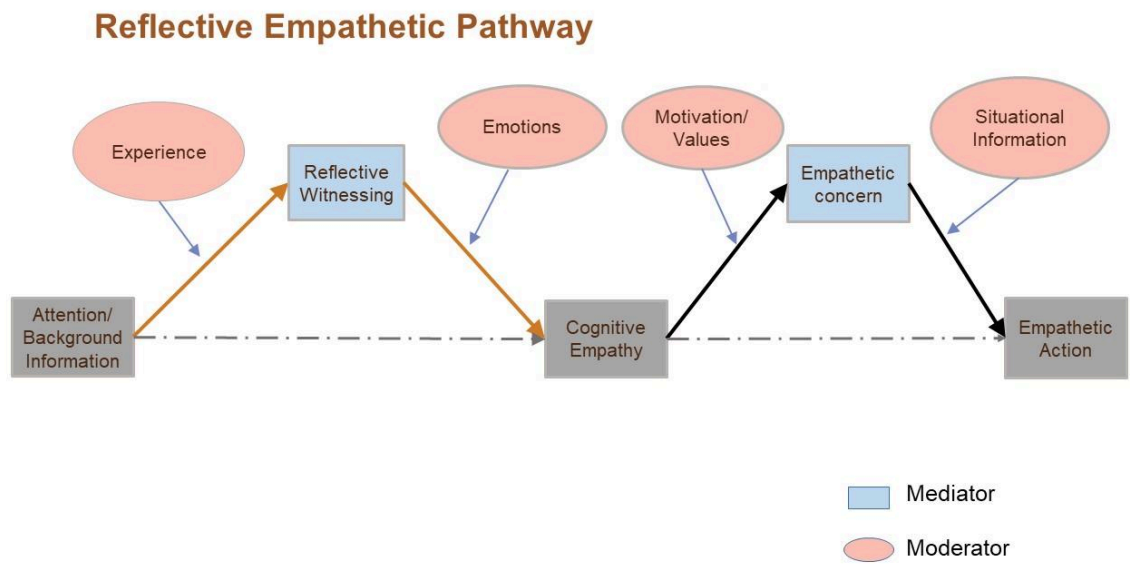

Figure 2. The reflective empathetic pathway built on cognitive empathy, with all identified mediators and moderators. The gray rectangles represent the factors from the prior empathetic framework. The blue rectangles represent the mediators identified in this study [reflective witnessing, and empathetic concern]. The pink ovals represent the moderators identified in this study [emotions, motivations/value, situational information] along each pathway. The orange arrows indicate this is a reflective pathway that requires intention and cognitive effort. The black arrows represent the pathway after the development of empathy, leading towards empathetic actions. The arrows are black because we did not find any differences in this pathway between cognitive and affective empathy. The dashed gray line represents the prior simplified empathetic pathway.

We explore cognitive empathy through the experiences during the COVID-19 pandemic of Dylan as they recount supporting a neurodivergent student with ADHD and depression. Dylan begins,

*"There was COVID, and there were more mysteries. I had a graduate student who was suffering, who [had a disability] and they were really suffering from being alone at that time. I...I'll ask them to write a plan of what they're going to do, and then [I] figured out that this student was having a hard time even feeding themselves. So I reached out to the department chair and asked them to intervene and do something like we could send the students back home. I noticed that on campus [they were] alone and it was, and we couldn't find the student for a while. So we had to try to zoom with them. It was a couple of stressful days. But as soon as they managed to get home, I met with them over Zoom, like monthly, to make sure that they're doing fine."*

Here we see Dylan noticing the student's struggles to submit a work plan and, by paying closer attention, learning of their feelings of isolation and struggles to navigate day to day life. Importantly, Dylan knows and learns important background and contextual information, such as the student's neuro-divergence and 'their hard time feeding themselves', implying that they struggled to access food and groceries during COVID. The importance of background information is seen here (and in other interviews) as a crucial part of attention/noticing, providing context necessary to interpret experiences. Here, we also see that obtaining background information is not a passive task, but rather an iteratively active process that often involves direct communication with students or others. Dylan subsequently performs an empathetic action by reaching out to the department chair to arrange for the student to go home and holding monthly Zoom check-ins.

An identified mediator is ***reflective witnessing***, a critical analysis of self, others, and the social phenomena at play. Reflective witnessing involves active listening, challenging narratives and assumptions and, importantly, *self-differentiation* in acknowledging one's own biases and experiences as different from another's. We see this in Dylan's processing of how they were feeling and what they were thinking about during this process ("It was a couple of stressful days"). Dylan continues to express feelings of fear and worry for the student, thinking deeply about the student's situation beyond the research at hand. They consider the student's health, stipend, accommodations to get home and prioritize the student's health over research and implicitly acknowledge the mental load by recognizing the relief felt once the student was safe.

*"I was mostly feeling like, we have to solve this problem as quickly as possible, because it might be a danger to the students' health [which is] the priority… It was very stressful for everybody. My concern was that the student had to get on a flight and go home. Once everything was arranged, and when they got home and they were healthy, that was a relief. Yeah. And I think for the rest of the summer, I was just fine because the student was with their parents."*

  ***Reflective witnessing*** is an explicitly cognitive process, allowing one to intellectually understand what another is experiencing. If one has had a similar experience, then these *shared experiences* can act as moderators, making it easier to engage in the process of reflective witnessing, although having shared or adjacent experiences is not essential for cognitive empathy. Dylan's feelings of stress, worry, and relief illustrate how *emotions* can act as moderators, facilitating the development of cognitive empathy. Empathy, whether cognitive or affective, is not complete without action. **Empathetic action** is the result of understanding and feeling another's emotions as tangible efforts to alleviate their suffering. It's the "doing" part of empathy, where we move beyond simply acknowledging another's experience and actively seek to support them. This spectrum of action can range from offering a listening ear to tackling larger systemic issues. However, the crucial element is the genuine desire to help, fueled by both cognitive and/or emotional understanding. Communication is the cornerstone of effective empathetic action. What one person finds helpful might not be what the other needs. Open and honest communication allows one to tailor their actions to the specific situation and ensure they truly benefit the other person.

      Dylan also indicated that they valued the student's health and thought of it as a priority. Dylan's *values* motivated them to develop a strong *empathetic concern* for the student and allowed them to take action. Dylan's *situational information* allowed them to reach out to their department chair and express their concern and advocate for their student. **Empathetic concern** goes beyond mere sympathy. It's a complex emotional state that fuels a deep-seated desire to improve another's well-being.  Unlike sadness or pity, it compels us to act. Cultivating this concern requires self-reflection – understanding how our values and motivations influence the agency one feels and determines the kind of help we offer, thus motivation and values are an important moderator in this process.

Dylan continues to display empathetic concern for the student which translates into sustained action:
*" For me, it [staying connected with the student over summer] was more of a drag after the student got home, it was more of a drag because we didn't really have any [research] we did together. We [Dylan and department head] decided not to give them any work because my work was experimental. So it was just, it [having weekly zoom meetings] was a personal thing I was doing, it was not really related to work.  The student who was neurodivergent and experiencing depression didn't have anything to talk about. So I had to come up with a list of things I'm going to ask the student. It was..., but I can't say it was a drag, it was hard for me. Yeah. The student was — I can't read their emotions, I couldn't really read their emotion. But it was obviously they were doing much better because when the student was on campus, they always had like, very untidy hair, and it was obvious that they're not really taking care of themselves. But when I saw them zooming from home, it was always nice and clean clothes. And so it looked like they were doing well, and apparently they actually were looking forward to these meetings. So it gave them a sense of I'm still a student. And I have things to do, although we got nothing done that summer, but it was not bad."*

Here we see a toll on Dylan as they learn to meaningfully communicate with the student despite not fully emotionally understanding them (indicating a cognitive, not affective empathy). We also see the development of Dylan's *noticing* as they comment on mannerisms and appearance and recognize the impact the meetings have on the student. Once the summer ends, Dylan continues to navigate this situation through an empathetic lens.

*"We [Dylan, student, and department head] had time to reflect on how the student was performing. And if they can be successful, successful in my research lab. I consulted with faculty in their own department who were program directors, and they decided that the student should leave my lab eventually. It was mostly due to the nature of my work and the personality of the student, it would have been forced for the student to actually stay in an experimental lab and not be able to really flourish. Given the circumstances, I think it was the best thing for the student. And also, they [program director] wanted them to work with somebody who was older, or older than me and more experienced with different personalities. So there are two sides of the story, you want the student to be successful, and you want the faculty not to only, like, be drained by one student. So the faculty who picked up the student had a much larger team, and they could incorporate the student with other students."*

Dylan's narrative outlines the cognitive empathy pathway from noticing to action. We indicate in Table 1 specific experiences articulated by Dylan that pertain to the pathway, mediators and moderators. Table 1 represents the themes (factors) on the left column, with details of each of these factors identified by the narrative in the right column. These bulletin points stem from codes but build on them to be more specific to this narrative.

| Attention/Noticing/ Background Information | <ul><li>Difficulty with research</li><li>Living alone on campus</li><li>Neurodivergent student with ADHD and depression</li><li>COVID-19 consequences</li><li>Student struggling w/everyday tasks</li></ul> |
|---|---|
| Reflective witnessing | <ul><li>Considers how background information connects to situation</li><li>Tries to put self in place of student</li></ul> |
| Emotions | <ul><li>Articulates feelings of stress, worry, fear and relief</li></ul> |
| Empathetic concern | <ul><li>Concern for students well-being</li></ul> |
| Motivation/Values | <ul><li>Values student wellbeing</li><li>Motivated by sense of responsibility</li></ul> |
| Situational information | <ul><li>Relatively new faculty member</li><li>Small research group</li><li>Experimental lab</li></ul> |
| Empathetic action | <ul><li>Arranges w/ Dept. Chair for student to go home, get stipend money</li><li>Continues to check-in with student</li></ul> |

| | ● Seeks input on how to communicate with neurodiverse students |
|---|---|

Table 1. The Cognitive empathetic pathways filled in through Dylan's experience supporting a student with ADHD who was struggling during the pandemic. The left column represents the themes (factors) identified while the right column outlines and expands upon the codes (details) identified for this narrative.

Figure 2 integrates the mediators and moderators into this pathway. We note that the depicted pathway is iterative, a complexity not fully captured in a static representation. Dylan's narrative can be segmented into three distinct iterative cycles, each initiated by a specific action. The first begins with the identification of student challenges and progresses through cognitive empathy to culminate in the communication with the department chair. The second begins with Dylan recognizing through reflective witnessing a deficit in their knowledge/practices, assessing the student's financial situation, and, as a result, finding a way to continue the student stipend, participating in workshops on supporting neurodivergent students, and continuing to meet monthly with the student. The last cycle involves Dylan's actions in the post COVID-19 semester. Dylan communicates with the student and others and, drawing on situational information and grounded in the student's well-being, concludes that the student would do better in a larger group.

### B. Affective empathy pathway

Affective empathy is an intuitive process rooted in shared experiences. Faculty members interviewed largely agreed that affective empathy is easier than cognitive empathy and often characterized it as a reflexive, automatic response triggered by personal familiarity. Although most faculty members suggested that affective empathy comes more naturally to them, we found it challenging to elicit many examples of affective empathy. This difficulty may stem from faculty members' tendency to separate their personal and professional identities, impacting their responses. Additionally, our protocol did not specifically request examples of affective empathy until the fourth round, which might indicate that faculty are more primed to think about cognitive empathy when asked for general classroom examples. Another possibility is that the interviews were quite personal, which may have made many participants uncomfortable sharing examples related to affective empathy. Figure 3 graphically illustrates the affective empathy pathway including the identified mediators and moderators. Through this pathway one develops affective empathy, can recognize others emotions and often mirrors them as well. Once this understanding is developed one can engage or disengage towards potential actions. Having a shared or adjacent experience is crucial in developing an emotional understanding of the other person and faculty members' own emotions regarding said experience can hinder or strengthen the path towards affective empathy.

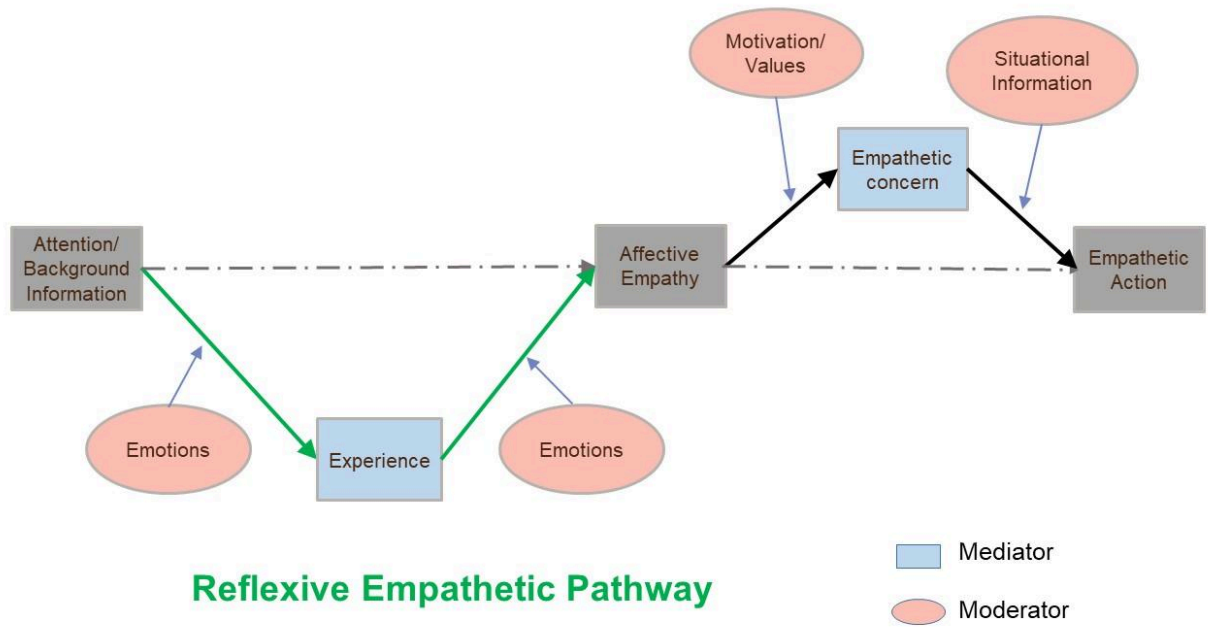

Figure 3. The reflexive empathetic pathway built on affective empathy, with all identified mediators and moderators. The gray rectangles represent the factors from the prior empathetic framework. The blue rectangles represent the mediators identified in this study [experience, and empathetic concern]. The pink ovals represent the moderators identified in this study [emotions, motivations/value, situational information] along each pathway. The green arrows indicate this is a reflexive pathway that is often passive and intuitive.

One example of a story with affective empathy is that of PM as he notices a student struggling to understand the physics concepts and thinks back to his experience as a physics student. PM states,

*"So this goes back to that role as someone who's been through the whole educational process where I can talk to somebody who's having trouble in the classroom and acknowledge for them that maybe they just had a really bad exam. And I'm not gonna tell them not to feel bad about it, because it feels bad. I know, it feels bad, because I had bad exams. When I was in grad school, and I would call my mom sobbing about how I should just drop out or something like that. And I didn't drop out. I'm here with a physics PhD now. So I know it feels really, really bad right now. But it's not going to feel really, really bad, forever. You will not just survive this, you can recover from this as a student. And I'm saying that as someone who was a student who had the similar stresses. So there I'm talking from a voice of experience, trying to commiserate based on having been in a similar situation before and being older, and for the wiser now."*

Similar to cognitive empathy, *background information* is typically gathered through active processes such as communication with the person experiencing the situation or sometimes by consulting external sources like experts or individuals with similar experiences. The background information can then help activate the reflexive affective empathy. We notice PM using this information to compare to his own experiences, which allows him to connect emotionally with the student. Similar to this example, stories of physics faculty experiencing affective empathy relied on some form of familiarity with the experience or a personal connection to the person. Thus experience acts as a mediator between the information obtained by attention/noticing and affective empathyPM knows that the student did poorly on an exam and connects this back to when he was a student facing similar circumstances. Importantly, PM recalls his *emotional* reaction and emphasizes his emotional understanding of where the student is coming from. PM's decision to not just tell the student to "not feel bad" stems from his perception that this is not consistent with the emotional reaction in the moment, showcasing his ability to identify and perceive the student's emotions.

While physics faculty's own *emotions* can influence the relationship, it does not seem to be necessary in developing affective empathy, and so we consider it a moderator rather than mediator. These emotions may not necessarily mirror those of the other person and can even block the pathway to empathy, as in a case where both faculty and student go through the same experience but have different emotional responses. For example, Bella talks about noticing a gender barrier in her experimental physics lab, where a female student was struggling with being a minority. Bella explains:

*"There was one gender barrier. I had one female student [in my lab] and that was very hard for the student. I did not feel like it should be a problem, but it turned out to be complicated for the students. And my gender [as a female] didn't really help, because the students [and I] have pretty different personalities."*

Bella, who identifies as the same gender as the student [female], did not experience a similar emotional response when she was the only female in a previous research group, as a graduate student. As a result, she initially believed that being the only female in the current group wouldn't be a problem. However, Bella acknowledges that her own lived experiences and emotions may differ from the student's due to their different personalities. Through this example, emotions act as a moderator with the potential to affect the strength of the connection between background information, experience and affective empathy instead of acting as a hidden variable along that path. In this case, the difference in emotions experienced by Bella compared to her student, weakened the path towards affective empathy, making it harder for her to emotionally understand the student, despite the similarity in experience.

As with noticing, *shared experiences* can make it easier to connect and understand other's emotions, which is also shown in the example of Kali. Similar to PM, Kali leans on a shared experience through remembering their time as a student, to develop an affective understanding of a student. Kali states:

*"Yeah, I had a student, [Mario], who was a grad student here. And then he went to [English University]. And he's doing his master's in England, where he's doing really well for himself. So he was complaining that his roommates were making too much noise. I related to*

*every one of it because I was that same person, that sensitive, highly high-strung, sensitive person who could not stand footsteps above my head and could not stand cigarette smoke, I was that very person. So I relate to everything he was saying. So I guess it requires you to have been through some of those experiences."*

Kali's perception of Mario as having a similar personality, coupled with his recollection of a similar experience during his own graduate school days, demonstrates how his **affective empathy** for Mario is rooted in their shared experiences. On the other hand, PM did not talk about his own emotions when interacting with the students and so emotions are not mediating the pathway despite him experiencing affective empathy. This indicates that emotions can amplify the affective empathy connection but do not constitute a necessary intermediary step.

PM continues the story by explaining his actions when he notices a student is struggling.

*"[I say to the student,] 'So here is what we can do as far as granting extensions about things.' So thinking about the tools that I've got, expressing sympathy to people, hearing them out, directing them towards whatever resources might be able to remedy or offer a partial remedy. Like, here's what we can do as far as extensions go, here are resources you can go to on campus to talk about stuff."*

PM shows that he is prepared with an outline of actions he can take and acknowledges the communicative nature of empathy and the value of listening to the student. These types of actions are mediated through his and the student's *situational information*. Later PM talks about having empathy for other faculty members which shows how the situational information impacts his empathy and **empathetic actions**.

*"For amongst faculty members, like I'm on the younger side of the faculty. So insofar as I don't want to say talking down to students, because that sounds condescending, right, like, that's not the right word for it. But that is that idea of me being able to come down from my position as someone who was there before, and so has broader perspective for it… versus that's never going to be the dynamic amongst me and peers, because we are peers, as opposed to somebody who's coming from a more experienced context...So I can say when they're [colleagues] going through something rough or like when somebody was a jerk to them in a meeting or whatever, I can tell them, 'That wasn't okay, if you want me to talk to department about it I can back you up on this.' But I'm not speaking from experience, as I would be when I'm speaking to students."*

PM draws on power dynamics between speaking with peers or students as part of the situational information that mediates his actions. PM is empathetically concerned for the other in both these situations but moves towards actions differently. We also see that PM's recognition of the power differential between him and students motivates him to give advice and help as he feels a sense of responsibility as someone who has succeeded through that path. PM's affective empathy relied on him thinking back to his past experience and drawing on those emotions. Thus, we see the same pathway from empathy to empathetic actions as shown in Dylan's example. PM *values* helping students build resilience and connecting them with useful resources/supports. This is especially crucial in situations like faculty-student interactions, where roles and responsibilities can differ.

| Attention/Noticing/ Background Information | <ul><li>Grade on test</li><li>Student values grades</li></ul> |
|---|---|
| Experience | <ul><li>Recalls difficulties they faced during graduate school courses</li></ul> |
| Empathetic concern | <ul><li>Concerned about student dropping out</li></ul> |
| Motivation/Values | <ul><li>Values open communication in classroom setting</li><li>Motivated by being a source of support for students</li></ul> |
| Situational information | <ul><li>Factors positionality in relation to situation</li></ul> |
| Empathetic action | <ul><li>Proves space/time for student to voice their emotions/concerns</li><li>Actively listens</li><li>Utilizes expertise to customize a plan of action specific to student</li></ul> |

Table 2.  Table of affective empathetic pathway filled in through PM's experience trying to support a student who did bad on an exam through reflecting on his own similar experience

Table 2 provides an illustration of how PM's experiences align with the affective empathy framework. PM uses his similar experience to emotionally understand what the student is feeling. The actions he decides to take depend on communicating with the student and finding what will be most helpful to them. PM values open communication and wants students to be able to approach him when they need help. He sees himself as someone who has passed the trials and tribulations of studying physics so can draw on his experience to help students navigate their own.

C. The combined pathway

Figure 4 integrates the two identified pathways for developing empathy. Both cognitive and affective empathy begin with paying attention to and noticing students, followed by gathering background information. This process is complex, influenced by various factors including what faculty choose to focus on and their own identities, as familiarity often makes certain aspects more noticeable. For simplicity, we group these influences into a single factor in this paper, while acknowledging their inherent complexity.

Experience and emotions play roles in developing both cognitive and affective empathy. Experience is crucial for affective empathy but not for cognitive empathy, which is why it is represented as a moderator in the cognitive pathway. Both cognitive and affective processes can evoke emotions in physics faculty, which may or may not align with the students' feelings. Therefore, developing cognitive empathy does not preclude experiencing emotions, as learning new information often elicits feelings, though these may not always mirror the students' emotions. Once cognitive or affective understanding is achieved, our data showed no significant differences in how faculty decided on the actions they would take.

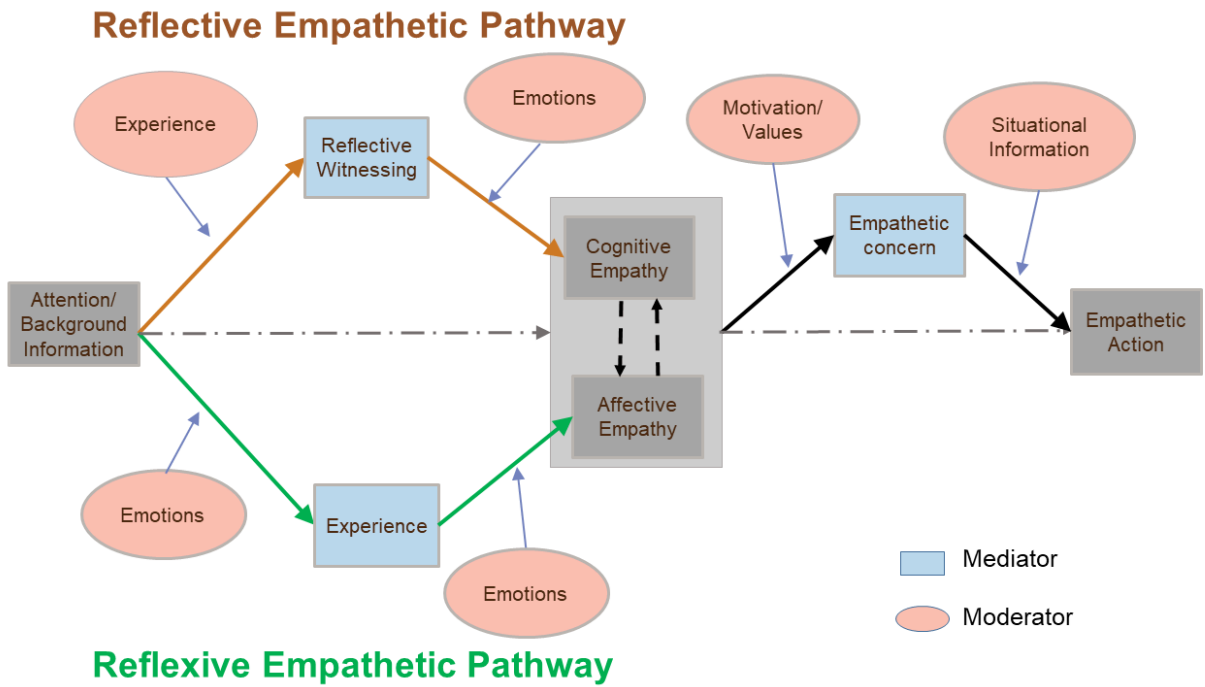

Figure 4. Complete framework for empathetic pathways in physics faculty. The gray boxes shown indicate the framework developed in prior work (Merril et al., 2024), the dashed gray line also represents the general pathway from paying attention/ noticing towards cognitive or affective empathy then one's empathetic actions. This image illustrates two main pathways; the orange pathway represents the reflective route towards cognitive empathy. This pathway is mediated by reflective witnessing; the practice of actively listening to another person's experiences without judgment or interruption, and then intentionally reflecting on the sociocultural dimensions of what was said, and moderated by experiences and emotions. The green route represents a more reflexive path towards affective empathy that is mediated by having a shared or adjacent experience, and moderated by one's emotions. We identified that the path towards empathetic action is mediated by empathetic concern; which encompasses what the faculty member thinks of as the ideal action, and moderated by physics faculty members' motivation and values while developing empathetic concern and situational information when deciding on how to use their understanding towards reasonable actions.

V. Discussion

A. Mediators, Moderators and the empathetic framework

By integrating both pathways, we have constructed a comprehensive empathetic framework that expands upon previous theories. This model incorporates mediators and moderators to elucidate the intricate process depicted in Figure 4. The reflective pathway is developed when employing cognitive empathy, while the path towards affective empathy is more reflective. Both cognitive and affective empathy are shown to take the same pathway towards empathetic actions. The actions physics faculty take are highly dependent on the individual which we see through moderators like one's personal motivation and values as well as the situational information. We chose to use dotted lines to indicate the prior framework elements as illustrated in Merrill et al. (2024). Yu and Chou (2018) hypothesized a direct link between cognitive and affective empathy however our data does not include an example from physics faculty that have moved between the two constructs. It is possible that faculty must put some effort into learning to notice things about themselves or the student to prompt them to follow the other pathway. For this reason we chose to use dotted lines between the two empathetic constructs as well. Instead of using "prosocial behavior" as defined by Yu and Chou (2018) and "prosocial concern" as described by Zaki and Ochsner (2012), we opted for the term "empathetic action" to emphasize actions taken (or not taken) based on one's empathetic understanding of others. The term "prosocial" was limiting because it suggested that actions are always intended to benefit others, which is not always the case. Empathetic actions are also influenced by personal implications and are constrained by the faculty members' capacity and their perceptions of their responsibilities.

Affective empathy, characterized by shared emotions, acts as a quick and direct pathway. Cognitive empathy, on the other hand, requires a more deliberate effort. We can see from diagram 4 that for both, background information is an important part of the factor attention/noticing which is necessary before the empathetic cycle even begins. Reflective witnessing, a critical process involving active listening, challenging assumptions, and self-awareness, is identified as a *mediator* in the cognitive empathy pathway. It requires intellectual understanding of another's experiences which may be facilitated by shared experiences. Experiences and emotions *moderate* the cognitive empathy pathway, facilitating reflective witnessing and intellectual understanding but not being essential for these processes, as demonstrated by Dylan's understanding of the difficulties faced by their autistic student without having a similar experience or facing similar emotions. In contrast, experience acts as a mediator in the affective empathy pathway as it is foundational for developing emotional understanding. Affective empathy is a reflexive process, and physics faculty lean on their similar experiences to develop a quick emotional understanding of others. PM and Kali both think back to their past experiences when navigating their students' struggles, be it something more generic like doing poorly on a test or more specific like being bothered by dorm-mates. Once faculty reach an understanding of the others situations, whether cognitive or affective, than they have the opportunity to decide on what actions to take,

Empathetic concern, situational information, and motivations and values are identified as mediators and moderators in the pathway towards empathetic action. *Empathetic concern* acts as an internal compass, guiding one towards ideal responses and actions. Physics faculty need to be internally *motivated* to move beyond understanding another's situation and take concrete steps to help. Empathetic concern thus encompasses what the faculty member thinks of as the ideal action. However, actions are also influenced by external factors like positionality (e.g., career level) and external rules, which we termed as *situational information*. Thus, the situational information moderates the actions taken. Despite the constraints imposed by various factors, faculty members retain the autonomy to navigate these challenges and select meaningful actions within their purview. This is evident in PM's discussion of his interactions with students and colleagues, where he reflects on the power dynamics inherent in these different contexts. In simpler terms, *empathetic concern* is the "what" – the thoughts about possible actions to help. *Empathetic actions* are the physical manifestation of that concern – the "how" we translate those thoughts into tangible support. Empathetic actions lie on a spectrum and can range from doing nothing, communicating with that individual or others relative to them, obtaining resources, lending a listening ear or more engaged steps to create change. There is no right or wrong type of empathetic action, and the steps taken depend on the context and the individuals involved.

### B. Addressing Bias in Empathy

Empathy is a contingent construct when connecting with others and can have both positive and negative impacts. Critics (Zhou, 2022) have expressed concerns about an *empathy bias* in which faculty connect more easily with those like themselves. This can lead to unequal treatment, as individuals are more likely to help those they perceive as similar to themselves (the "in-group") (Hein et al., 2010). This bias is particularly dangerous in physics given the mostly homogeneous nature of the field and has been confirmed (Dancy & Hodari, 2023). The empathetic pathway suggests possible actions to address this, in particular taking an active and reflexive approach to empathy as defined within reflective witnessing. Promoting and utilizing cognitive empathy, while recognizing that affective empathy may not be possible, can help faculty develop the critical understanding of students and colleagues and decide on meaningful actions. Because cognitive empathy is not automatic, it calls for intentional effort and time on the part of faculty and departments to explicitly mitigate the empathy bias. For example, we see Dylan developing a plan for empathetic action through intentional reflection and active knowledge-seeking. Their deliberate choices to participate in a workshop, maintain regular meetings, and consider long-term solutions were grounded in cognitive empathy and significantly impacted the student. This demonstrates that while affective empathy may be more instinctive, cognitive empathy can be equally effective for physics faculty. Our notion of reflective witnessing compels individuals to consider others' perspectives, differentiate themselves while acknowledging the other's validity, and actively listen and learn.

### C. Empathy as a co-constructed construct

Empathy is a reciprocal process that relies on trust building and vulnerability even in educational settings. It requires active participation from both faculty and students at various levels: individual interactions, research labs, classrooms, and even departmental collaborations. Traditional models often view empathy as an individual skill, neglecting the importance of communication in fostering this connection. However, we find many reminders that empathy is a dynamic process, evolving and adapting to the social context of the interaction. Similar to learning theories, we acknowledge the different ways people experience and express empathy. This "co-creation" of empathy moves beyond a simple "learn and give" dynamic. In the faculty-student relationship, this co-creation is even more crucial due to the inherent power dynamic. The model presented in this paper focuses on the perspective of an individual faculty member, and as a result, it does not capture the reciprocal aspect of empathy. Future research will aim to explore the co-creation of empathy in greater depth, emphasizing the role of communication in this process. Communication intersects with all components of the current model. For instance, acquiring background information frequently involves conversing with others, and listening to a student's experiences can prompt memories of similar situations. Communication can also be considered a form of empathetic action, as demonstrated by Dylan seeking advice from more experienced colleagues or maintaining ongoing dialogues with students. Similarly, PM listens to a student's emotional response to poor exam performance and chooses responses that reflect his understanding of their feelings. The process of reflective witnessing also is grounded in listening to learn from the person, but can also incorporate learning through other methods like reading or watching videos on a topic and involves an internal reflection about self and the general socio-political context. By framing empathy as a collaborative construct, we can potentially reduce compassion fatigue for faculty and open new communication channels. This fosters deeper understanding and leads to more effective actions that benefit both educators and students.

### D. Limitations and further work

Several limitations may affect this study's generalizability and findings. Focusing solely on one-on-one interactions does not address empathy within group dynamics, such as how it manifests in classroom settings or faculty interactions with student groups. The model presented was also developed from the perspective of the physics faculty member, thus we have no way of determining what students might have thought about each interaction or experience. Faculty might view their actions as empathetic, and describe them as such, but students could have a different perspective. This study was conducted at an R2 institution with a longstanding focus on teaching, which may limit the generalizability of the findings. Physics faculty at R1 institutions or community colleges might not align with the current model due to differing institutional priorities and contexts. To address these limitations, future research should aim to include a larger and more diverse sample that spans various demographics, faculty positions, and types of institutions. Additionally, incorporating both individual and group contexts would provide a more comprehensive understanding of how empathy is developed and enacted across different educational settings. The inherent positionality of the researchers as interviewers, along with some interviewers having existing connections to participants, is noted with both potential positive (participants may feel more comfortable with those they know) and negative (they might hesitate to share details) consequences.

## VI. Conclusion

We have presented a framework for an empathetic pathway in physics faculty, grounded in interviews and building upon existing research. Focusing on physics faculty, we identified two primary pathways utilized in developing empathy and determining appropriate actions. The framework expands previous models by incorporating mediating factors within both cognitive and affective routes of empathy. The framework encourages educators to move beyond simply acknowledging differences. In particular, empathy is the common theme through many efforts to increase participation and foster an inclusive environment. Culturally relevant pedagogies, accessibility, and universal design all rest on a meaningful understanding of what best supports individual students. This understanding, we argue, begins with empathy. Explicitly paying attention to how one empathizes can help faculty cultivate a classroom culture that celebrates diversity and leverages the unique perspectives of all students. This inclusive environment can have a profound impact on student engagement and success, particularly for those from underrepresented backgrounds in physics.

## VII. Acknowledgements

We extend our sincere gratitude to Rachel Merrill for her crucial contributions to data collection and for initiating this project. Our thanks also go to Christian Solorio and Ben Zwickl for their invaluable feedback on the manuscript, which greatly enhanced its quality. This work would not be possible without the participants who took the time to speak with us. We are grateful for allowing us to have vulnerable conversations and the emotional labor incorporated in that.We are grateful to CATSLE for their continuous input throughout various stages of the project, particularly their insightful advice on visualizations. This work was funded by National Science Foundation Grants No. DGE-2222337.